\documentclass[aps,prl,twocolumn,superscriptaddress,10pt]{revtex4-1}

\usepackage{graphicx}
\usepackage{dcolumn}
\usepackage{bm}
\usepackage[utf8]{inputenc}
\usepackage[pdfborder={0 0 0},hypertexnames=false]{hyperref}
\usepackage[all]{hypcap}

\begin{document}

\title{Quantum-limited measurements of optical signals from a geostationary satellite}

\date{\today}

\author{Kevin Günthner}
\author{Imran Khan}
\author{Dominique Elser}
\author{Birgit Stiller} \thanks{Now at the University of Sydney, School of Physics, NSW 2006, Australia}
\author{Ömer Bayraktar}
\author{Christian R. Müller}
\affiliation{Max Planck Institute for the Science of Light, Erlangen, Germany}
\affiliation{Institute of Optics, Information and Photonics, Friedrich-Alexander University Erlangen-Nürnberg (FAU), Erlangen, Germany}

\author{Karen Saucke}
\author{Daniel Tröndle}
\author{Frank Heine}
\author{Stefan Seel}
\author{Peter Greulich}
\author{Herwig Zech}
\affiliation{Tesat-Spacecom GmbH \& Co.~KG, Backnang, Germany}

\author{Björn Gütlich}
\author{Sabine Philipp-May}
\affiliation{German Aerospace Center (DLR), Space Administration, Bonn, Germany}

\author{Christoph Marquardt}
\email{Christoph.Marquardt@mpl.mpg.de}
\author{Gerd Leuchs}
\affiliation{Max Planck Institute for the Science of Light, Erlangen, Germany}
\affiliation{Institute of Optics, Information and Photonics, Friedrich-Alexander University Erlangen-Nürnberg (FAU), Erlangen, Germany}

\begin{abstract}
The measurement of quantum signals that traveled through long distances is of fundamental and technological interest. 
We present quantum-limited coherent measurements of optical signals, sent from a satellite in geostationary Earth orbit to an optical ground station. We bound the excess noise that the quantum states could have acquired after having propagated 38\,600 km through Earth's gravitational potential as well as its turbulent atmosphere. Our results indicate that quantum communication is feasible in principle in such a scenario, highlighting the possibility of a global quantum key distribution network for secure communication.
\end{abstract}

\maketitle

Quantum mechanics has successfully undergone a number of fundamental experimental tests since its development~\cite{Pan2000, Aspect2015, Normile2016}. Still some aspects pose both a theoretical and an experimental challenge, such as the relation of quantum mechanics and gravity~\cite{Plaga2006, Ralph2014, Bruschi2014}. Quantum-limited measurements of quantum states traveling through long distances in outer space provide both an offer to test quantum mechanics under such extreme conditions and a prerequisite for its use in quantum technology~\cite{Rideout2012}. To this end satellite quantum communication~\cite{Rarity2002, Ursin2009, Merali2012, Elser2012, Bacsardi2013, Graham2015, Gruneisen2016, Biever2016} promises to provide the currently missing links for global quantum key distribution (QKD).
Important experiments in satellite quantum communication have been reported or are currently being devised and set up~\cite{Scheidl2013, Jennewein2013, Pan2014, Tang2016, Dequal2016, Carrasco-Casado2016, Gibney2016}.

This work presents and discusses quantum-limited measurements on optical signals sent from a GEO-stationary satellite. We report on the first bound of the possible influence of physical effects on the quantum states traveling through Earth's gravitational potential and evaluating its impact on quantum communication.

\begin{figure}
        \includegraphics[width=245px]{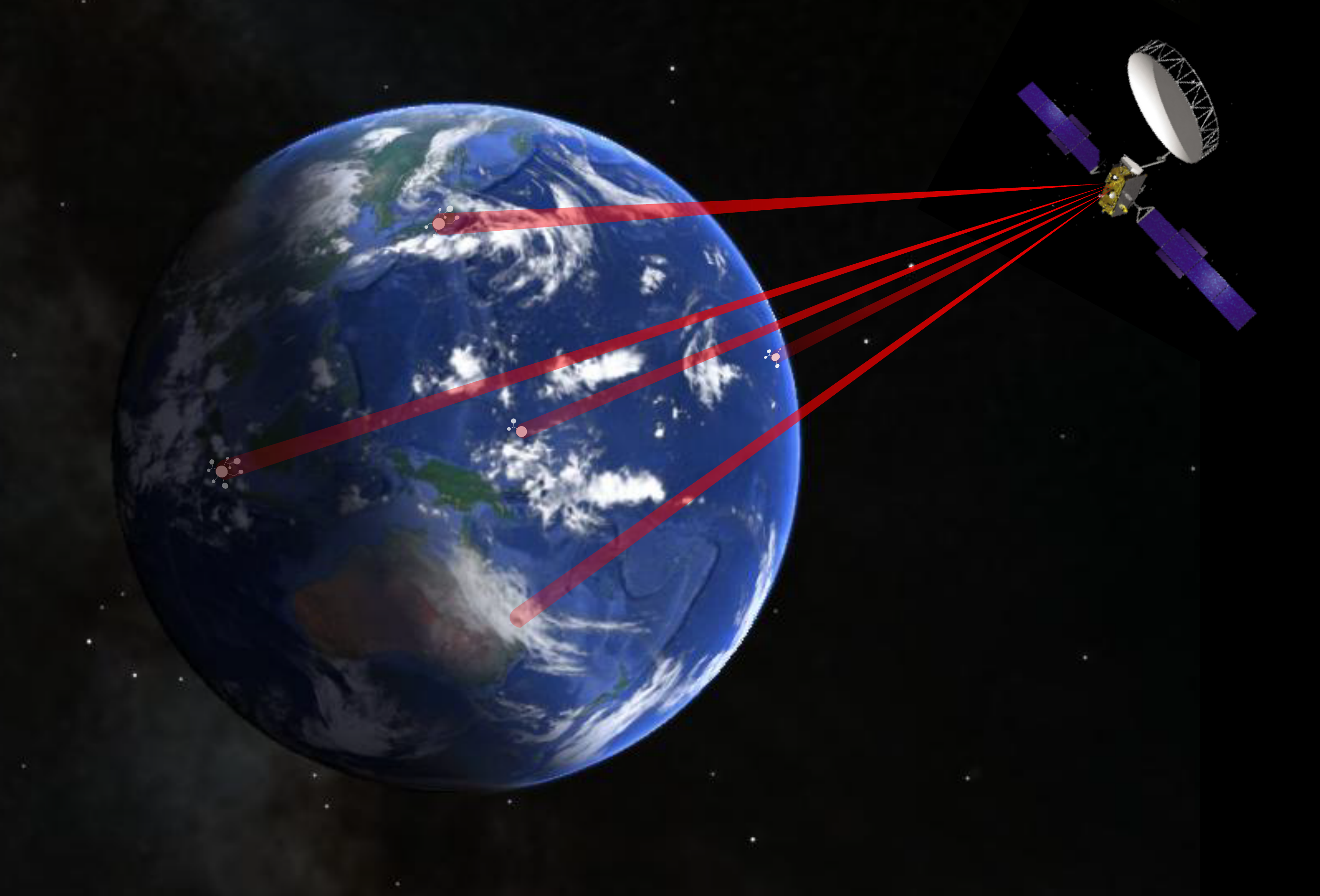}
        \caption{\label{fig1}Laser signals from geostationary Earth orbit travel through a large part of Earth's gravitational potential as well as through turbulent atmosphere. The successful characterization of quantum features under such conditions is a precondition for the implementation of a global quantum communication network using satellites. Metropolitan area quantum networks on ground would then be provided with the currently missing links among each other. (Picture of the Earth: Google, picture of the satellite: ESA)}
\end{figure}

Optical space-to-ground experiments for long-distance classical data communication have started in 1994~\cite{Arimoto1994} and have since been continued with successful demonstrations of optical ground links from low Earth orbit (LEO)~\cite{Fields2011}, geostationary Earth orbit (GEO)~\cite{Saucke2016} and the Moon~\cite{Boroson2014} (for a summary, see~\cite{Toyoshima2015}). In parallel, free space quantum communication has made its steps out of laboratories into real-world scenarios~\cite{Buttler1998, Schmitt-Manderbach2007, Ursin2007, Majumdar2015}. It has turned out that detecting field quadratures (continuous variables) is well suited to combat disturbances from atmospheric turbulence and stray light~\cite{Lorenz2004, Lorenz2006, Elser2009}. Using these methods, the first implementation of an intra-urban free space quantum link using quantum coherent detection has been reported~\cite{Peuntinger2014, Heim2014}. The advantage of stray light immunity applies as well to classical coherent satellite communication~\cite{Elser2015}.
The similarity between these classical and quantum technologies allows us to make use of the platform of a technologically mature Laser Communication Terminal (LCT)~\cite{Smutny2009, Heine2015, Zech2015} for future quantum communication (see Fig \ref{fig1}).

An important step on this way is to precisely characterize system and channel with respect to their quantum noise behavior. Coherent quantum communication employs encoding of quantum states in phase space and works at the limit of the Heisenberg uncertainty relation~\cite{Heisenberg1927}, but is susceptible to additional technical noise. Our task here is to characterize whether quantum coherence properties are preserved after propagation of quantum states over 38\,600 km, through a large part of gravitational potential of Earth and through turbulence of all atmospheric layers.

\begin{figure}
        \includegraphics[width=230px]{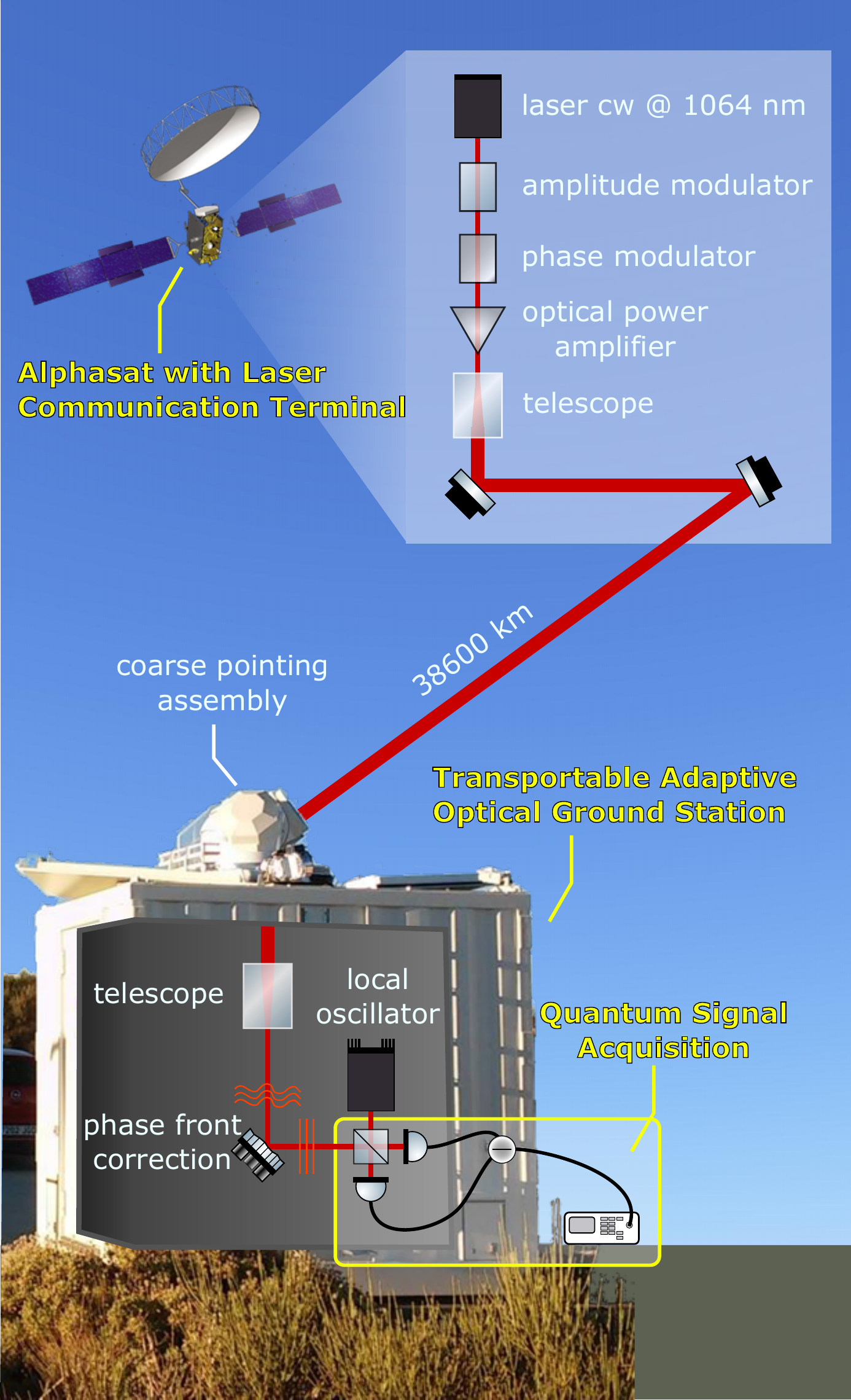}
        \caption{\label{fig2}Space-to-Ground Link Setup: A Laser Communication Terminal (LCT) on the Alphasat I-XL spacecraft in geostationary Earth orbit (GEO) links in continuous-wave (cw) mode to the Transportable Adaptive Optical Ground Station (TAOGS)~\cite{Fischer2015}, currently located at the Teide Observatory on Tenerife, Spain. The TAOGS is equipped with a Quantum Signal Acquisition System based on the homodyne principle where a weak quantum signal interferes with a local oscillator reference beam. By mode-matching the local oscillator to the signal, stray light is filtered out such that daylight causes no operational constraints. (Picture of Alphasat: ESA)}
\end{figure}

Our sender is a Laser Communication Terminal (LCT), flying on the GEO spacecraft Alphasat I-XL (see Fig. \ref{fig2}). The LCT was developed for classical coherent satellite communication (details on the LCT may be found in ~\cite{Smutny2009, Heine2015, Zech2015}). It is based on an Nd:YAG laser, operated at a wavelength of 1064 nm in continuous-wave mode. After passing through an amplitude modulator, a phase modulator at a frequency of 2.8 GHz imprints an alphabet of binary phase modulated coherent states $|\alpha\rangle$ and $|{-}\alpha\rangle$ on the light field~\cite{Glauber1963}. The same type of binary phase modulation can be used in quantum communication~\cite{Bennett1992,Heim2014}, together with quaternary phase modulation offering performance increase~\cite{Killoran2012,Heim2014}. Subsequently, the coherent quantum states are amplified and have thermal excess noise, defined as noise above the quantum uncertainty of the vacuum state. A periscope with diameter 13.5 cm precisely points the laser beam to the optical ground station. On ground, we use the Transportable Adaptive Optical Ground Station (TAOGS, see Fig. \ref{fig2}) for pointing, acquisition and tracking~\cite{Saucke2016} of the laser beam from space. An adaptive optics system corrects for phase front distortions in order to launch the beam into a single mode fiber (details on the TAOGS may be found in ~\cite{Fischer2015}). Our quantum acquisition system uses a homodyne detector~\cite{Yuen1978}, locked to the phase of the signal, to measure field variables in the continuous-variable (CV) regime~\cite{Ralph1999, Silberhorn2002}. This method allows for characterizing the quadratures of the light field with a precision at the quantum noise limit.

The total loss from the satellite output up to the receiving aperture was measured to be 69\,dB. 
This value incorporates the following sources of losses:
Most of the channel attenuation stems from the receiving aperture of 27\,cm being smaller than the spot size at the ground station. Loss through scattering and absorption at atmospheric particles is very low under good weather conditions and assumed here to be less than 2\,dB. Turbulence induced beam spread and scintillation result in a small loss of about 1\,dB. Besides that, there are losses through waveform aberrations and 
pointing errors.
Note, that beam wandering induced losses, which are important in shorter links~\cite{Vogel2016}, are small as the beam diameter is significantly larger than the receiving aperture.
Technical losses within the TAOGS are kept at a level of 16\,dB (including detector quantum efficiencies). Thus, in total the losses from the sending aperture to the detector add up to 85\,dB.

\begin{figure}
        \includegraphics[width=250px]{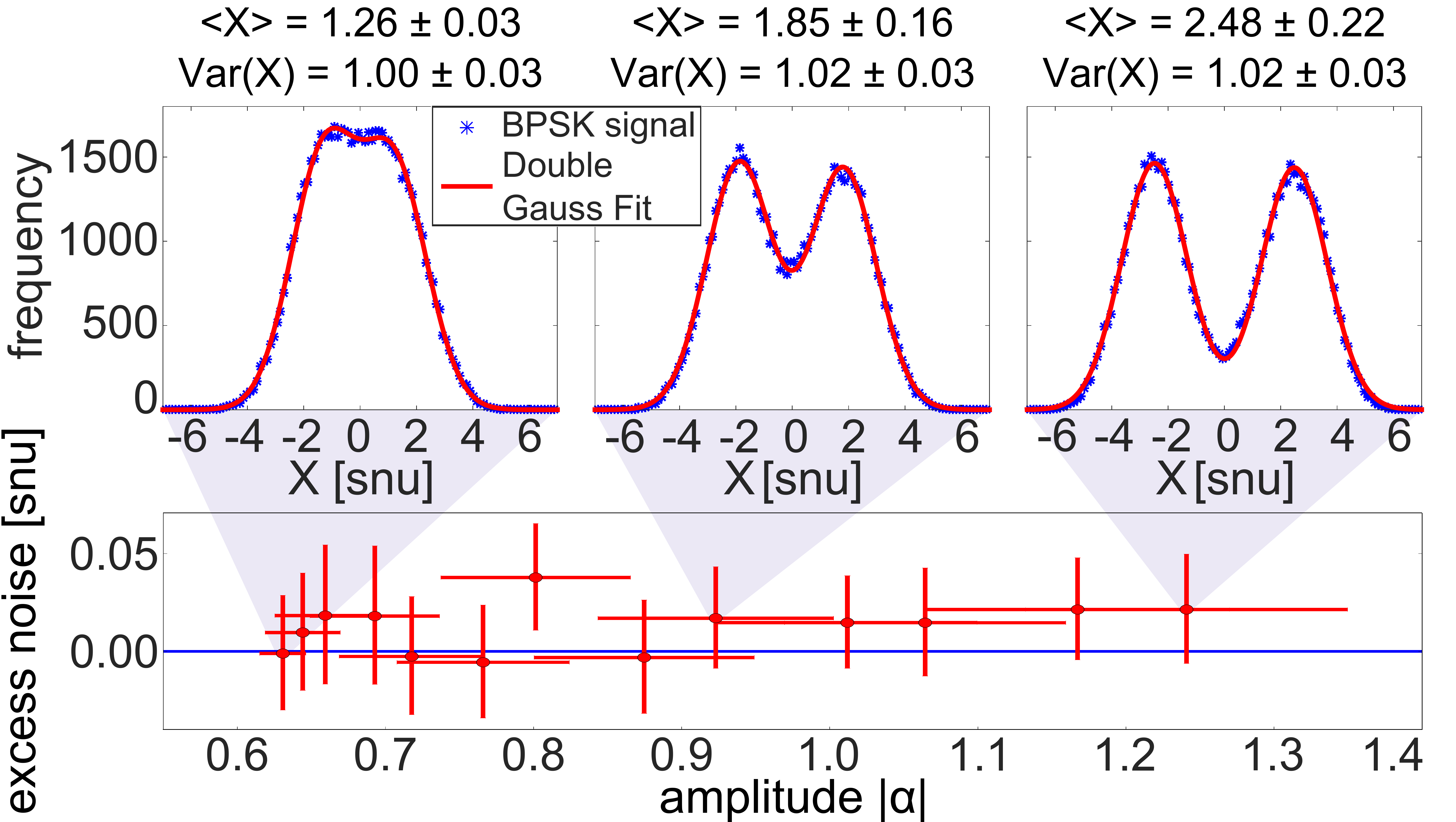}
        \caption{\label{fig3}Experimental results for the excess noise variance in units of the quantum uncertainty of the vacuum state. Data is shown for different detected signal amplitudes $|\alpha|$. In the upper row, three exemplary histograms ($|\alpha|=0.63, 0.92, 1.24$) illustrate the observed quadrature distribution along the $X$ quadrature. Each of the histograms contains about 70\,000 data points.}
\end{figure}

We focus our measurements and data analysis on the $X$ quadrature in which the signal is encoded~\footnote{The measurement of the $P$ quadrature will be subject of future studies.}. Details of the asynchronous sampling procedure are given in the Supplementary Material. The amplitude modulator in the LCT can sinusoidally vary the signal amplitudes, with slower frequency than the phase modulation. We acquire the signal states of varying amplitudes and sort them into 50 bins of quasi equal amplitude, each containing about 70\,000 measurement values. This method allows us to produce histograms of $X$ quadrature values for each bin (three of which are shown in Fig.~\ref{fig3}).

A binary phase modulation of coherent states produces a double Gaussian measurement distribution in a quantum-limited homodyne detection system. As we can see in Fig.~\ref{fig3}, our experimental data agree well with this prediction. Therefore, we can determine the amplitude and the variance of the received quantum states using a double Gaussian fit, assuming that the variances of the two states $|+\alpha\rangle$ and $|-\alpha\rangle$ are equal. We normalize our measured signal states to a reference measurement of the quantum vacuum state by closing the signal port of the homodyne receiver. We verify the linearity of the photodetectors and subtract the electronic noise variance from both signal variance and vacuum variance in all measurements (see Supplementary Material).

Dividing the signal variances by the vacuum variance yields the vacuum-normalized variances and allows to deduce the amplitudes of our signal states in the $X$ quadrature (see Fig.~\ref{fig3}). Our measured excess noise contains excess noise picked up during propagation through the atmosphere, as well as all potential technical noise in the ground station before detection. Nevertheless, we observe a signal variance of $1.01\pm0.03$ in units of the quantum uncertainty of the vacuum state and thus conclude that the detected signal states are almost quantum uncertainty limited coherent states. This means that our result is consistent with vanishingly small excess noise within the error bars.

Our measurement results show that coherence properties can be preserved even after very long propagation and through Earth's atmospheric layers. Using a low-noise phase-locking mechanism based on homodyne detection, quantum communication protocols using phase encoding become possible. The channel loss from GEO with an increased ground aperture of 1.5\,m would be reduced by 14\,dB from 69\,dB to 55\,dB. By improving the technical loss, discrete-variable QKD protocols, employing differential phase-shift keying or decoy states, operated from a GEO-stationary satellite, then become feasible~\cite{Inoue2002, Wang2005, Ma2005, Lo2005, Hwang2003, Shibata2014, Korzh2015, Bourgoin2015}. For continuous-variable (CV) QKD protocols with coherent detection, using Gaussian modulation~\cite{Jouguet2013} or Gaussian post-selection~\cite{Walk2013} an operation in low Earth orbit is conceivable.
Assuming an Alphasat-like LCT in a LEO orbit at a distance of 500\,km the channel losses of 69\,dB would be reduced to 31\,dB. Using a bigger receiving aperture of 1\,m could reduce the losses
by another 11\,dB.
Thus, the channel loss from a LEO satellite is estimated to be around 20\,dB, what is within the tolarable 
amount of loss given in~\cite{Walk2013}, even adding the typical amount of technical loss. 

These considerations may be extended to quadrature phase-shift keying (QPSK), bearing in mind that security proofs for CV discrete-modulation protocols still need to improve to support quantum channels with higher losses. Note, that future laser communication terminals for quantum communication require adjusted sending amplitudes and reduced internal excess noise by suitably multiplexing classical and quantum mode.

\begin{figure}
        \includegraphics[width=230px]{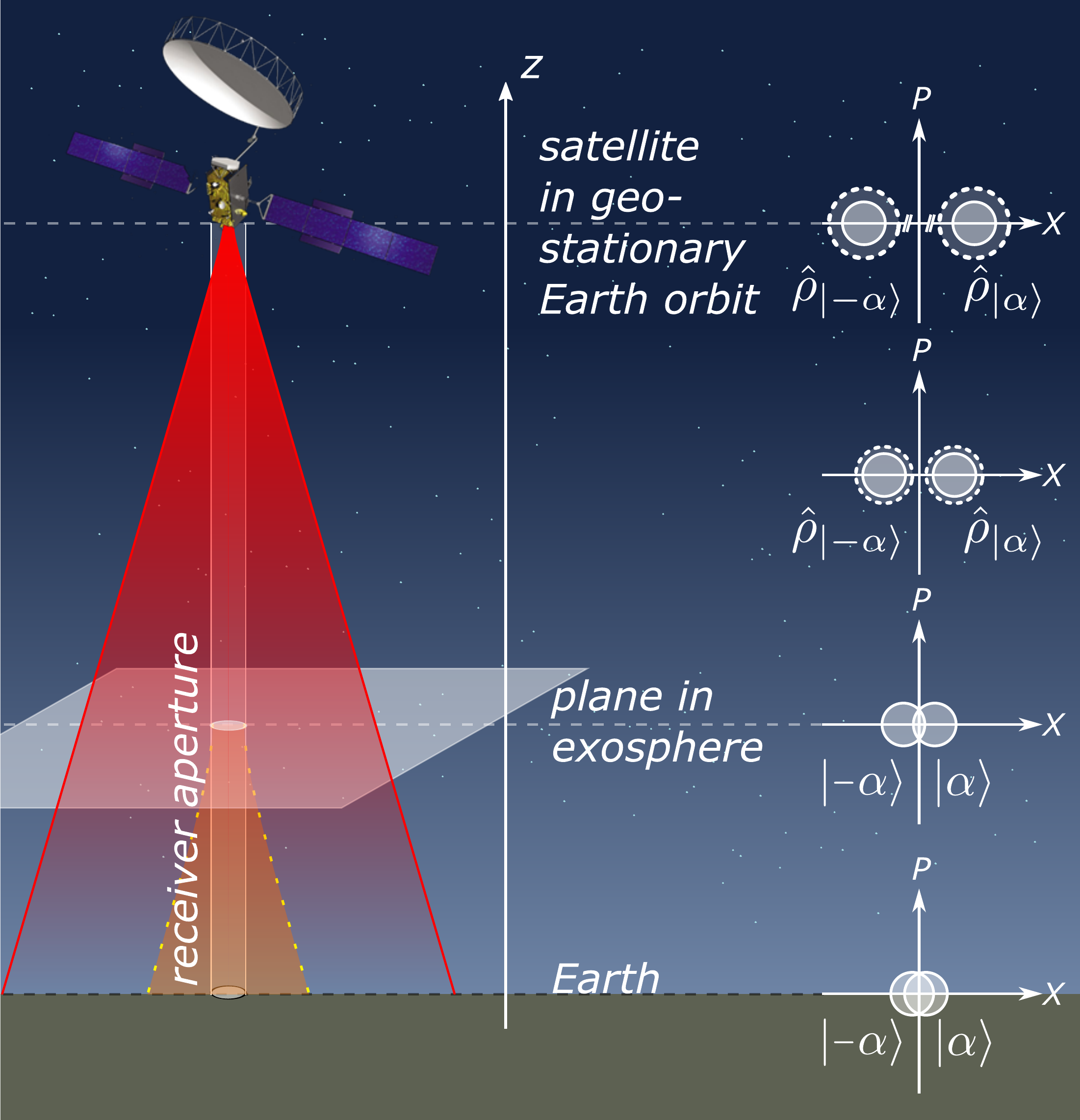}
        \caption{\label{fig4}The satellite in geostationary Earth orbit produces two phase-encoded thermal states. Due to the high channel attenuation, the thermal states converge to coherent quantum states on their way down to Earth. Therefore, a receiver at an altitude of 1\,000\,km above ground would detect nearly quantum uncertainty limited signals. At the same position, we can image a virtual aperture transmitting quantum uncertainty limited states. Using this model, we can estimate an upper bound for atmospheric influence of $0.8 \pm 2.4$ above the quantum uncertainty of the vacuum state (see Supplementary Material). (Picture of the satellite: ESA)}
\end{figure}

An interesting aspect is that our measurement data allows us to study the influence of the atmosphere on the propagation of quantum states. Initially, the states are prepared by the LCT with an excess noise of 33\,dB above the quantum uncertainty of the vacuum state. We assume that the propagation through the almost particle-free exosphere only has negligible influence on the quantum states. Imagine a virtual receiver with a receiving aperture of the size of the TAOGS aperture at an altitude of 1000 km above ground. This virtual aperture would observe an intensity reduced by a factor of 61\,dB, governed by the geometrical beam expansion (see Fig.~\ref{fig4}). Due to the same loss, excess noise would be reduced to merely 0.001 units above the quantum uncertainty of the vacuum state. Both excess noise picked up during propagation through the atmosphere and technical excess noise originating at the receiver on ground is visible in our measurement data. However, both the atmosphere and our receiver are not loss-free, which implies that higher excess noise than actually measured might have been present during propagation. To give a worst case bound for this value, we scale the measured excess noise up, from a value of $0.01\pm0.03$ by a factor of 16~dB (ground station losses) resulting in $0.4\pm1.2$. Additionally, we scale this value with atmospheric losses of 3\,dB and obtain an upper bound for the excess noise of $0.8 \pm 2.4$ (see Supplementary Material). Note that this is a conservative estimate for both the atmospheric and technical excess noise. Whereas excess noise originating during propagation has to be scaled up this is not the case for noise originating at the receiver on ground, lowering the bound significantly. According to our estimation both noise sources can not be distinguished. 

Using the same estimate, our quantum-limited measurements on ground bound the excess noise that could have been added during the whole propagation through the Earth's gravitational potential. Taking into account the overall loss of 85\,dB between satellite and detector on ground, we can give a first bound of any excess noise being below 65\,dB, well in agreement with the 33\,dB of thermal excess noise originating from the satellite (see section on amplifier noise behavior in the Supplementary Material).

Concerning the implementation of satellite QKD, we have presented phase space measurements of optical signals sent from geostationary Earth orbit (GEO) to an optical ground station. We have shown by measurement that quantum limited states arrive at the ground station despite the long propagation path including Earth's atmospheric layers. We have bound the overall excess noise that can degrade the quantum states in the satellite-ground link and the atmospheric layers. This work can be seen as the first step in developing quantum communication from GEO. While the currently existing hardware is optimized for classical coherent data communication, the technological proximity to coherent quantum communication opens up a fast and efficient way to develop a global quantum network. Moreover, beyond the strict QKD regime our results can be adopted to the overall context of security by physical means (see e.g.~\cite{Wilde2013}) as a powerful line of defense against future cyber-attacks~\cite{Endo2016}.

\paragraph{Acknowledgments:}
The Laser Communication Terminal (LCT) and the Transportable Adaptive Optical Ground Station (TAOGS) are supported by the German Aerospace Center (DLR), Space Administration, with funds from the Federal Ministry for Economic Affairs and Energy according to a decision of the German Federal Parliament.

We thank Bettina Heim, Christian Peuntinger, Nathan Killoran, Kaushik Seshadreesan, Radim Filip, Norbert Lütkenhaus, Christoph Pacher, Zoran Sodnik and Harald Hauschildt for fruitful discussions. 
We furthermore thank the European Space Agency (ESA) for hosting the Transportable Adaptive Optical Ground Station (TAOGS) next to the ESA Optical Ground Station (OGS) in Tenerife.

K.~G. and I.~K. contributed equally to this work.

\bibliographystyle{Science}
\bibliography{satqkdrefs}

\begin{thebibliography}{10}

\bibitem{Pan2000}
J.-W. Pan, D.~Bouwmeester, M.~Daniell, H.~Weinfurter, A.~Zeilinger, {\it
  Nature\/} {\bf 403}, 515 (2000).

\bibitem{Aspect2015}
A.~Aspect, {\it Physics\/} {\bf 8}, 123 (2015).

\bibitem{Normile2016}
D.~Normile, {\it Science News, Apr. 5, 2016\/}  (2016).

\bibitem{Plaga2006}
R.~Plaga, {\it Eur. Phys. J\/} {\bf 38}, 409 (2006).

\bibitem{Ralph2014}
T.~C. Ralph, J.~Pienaar, {\it New J. Phys.\/} {\bf 16}, 085008 (2014).

\bibitem{Bruschi2014}
D.~E. Bruschi, T.~C. Ralph, I.~Fuentes, T.~Jennewein, M.~Razavi, {\it Phys.
  Rev. D\/} {\bf 90}, 045041 (2014).

\bibitem{Rideout2012}
D.~Rideout, {\it et~al.\/}, {\it Class. Quantum Grav.\/} {\bf 29}, 224011
  (2012).

\bibitem{Rarity2002}
J.~G. Rarity, P.~R. Tapster, P.~M. Gorman, P.~Knight, {\it New J. Phys.\/} {\bf
  4}, 82 (2002).

\bibitem{Ursin2009}
R.~Ursin, {\it et~al.\/}, {\it Europhysics News\/} {\bf 40}, 26 (2009).

\bibitem{Merali2012}
Z.~Merali, {\it Nature\/} {\bf 492}, 22 (2012).

\bibitem{Elser2012}
D.~Elser, {\it et~al.\/}, {\it Proc. ICSOS\/} (IEEE, 2012).

\bibitem{Bacsardi2013}
L.~Bacsardi, {\it IEEE Commun. Mag.\/} {\bf 51}, 50 (2013).

\bibitem{Graham2015}
T.~Graham, {\it et~al.\/}, {\it Proc. ICSOS\/} (IEEE, 2015).

\bibitem{Gruneisen2016}
M.~T. Gruneisen, {\it et~al.\/}, {\it Opt. Eng.\/} {\bf 55}, 026104 (2016).

\bibitem{Biever2016}
C.~Biever, {\it Nature News, 13 January 2016\/}  (2016).

\bibitem{Scheidl2013}
T.~Scheidl, E.~Wille, R.~Ursin, {\it New J. Phys.\/} {\bf 15}, 043008 (2013).

\bibitem{Jennewein2013}
T.~Jennewein, B.~Higgins, {\it Phys. World\/} {\bf 26}, 52 (2013).

\bibitem{Pan2014}
J.-W. Pan, {\it Chinese Journal of Space Science\/} {\bf 34}, 547 (2014).

\bibitem{Tang2016}
Z.~Tang, {\it et~al.\/}, {\it Phys. Rev. Applied\/} {\bf 5}, 054022 (2016).

\bibitem{Dequal2016}
D.~Dequal, {\it et~al.\/}, {\it Phys. Rev. A\/} {\bf 93}, 010301 (2016).

\bibitem{Carrasco-Casado2016}
A.~Carrasco-Casado, {\it et~al.\/}, {\it Opt. Express\/} {\bf 24}, 12254
  (2016).

\bibitem{Gibney2016}
E.~Gibney, {\it Nature\/} {\bf 535}, 478 (2016).

\bibitem{Arimoto1994}
Y.~Arimoto, {\it et~al.\/}, {\it Proc. SPIE 2381\/}, G.~S. Mecherle, ed.
  (1995), p. 151.

\bibitem{Fields2011}
R.~Fields, {\it et~al.\/}, {\it Proc. ICSOS\/}, E.~M. Carapezza, ed. (IEEE,
  2011), p.~44.

\bibitem{Saucke2016}
K.~Saucke, {\it et~al.\/}, {\it Proc. SPIE 9739\/}, H.~Hemmati, D.~M. Boroson,
  eds. (2016), p. 973906.

\bibitem{Boroson2014}
D.~M. Boroson, {\it et~al.\/}, {\it Proc. SPIE 8971\/}, H.~Hemmati, D.~M.
  Boroson, eds. (2014), p. 89710S.

\bibitem{Toyoshima2015}
M.~Toyoshima, {\it et~al.\/}, {\it Proc. ICSOS\/} (IEEE, 2015).

\bibitem{Buttler1998}
W.~T. Buttler, {\it et~al.\/}, {\it Phys. Rev. A\/} {\bf 57}, 2379 (1998).

\bibitem{Schmitt-Manderbach2007}
T.~Schmitt-Manderbach, {\it et~al.\/}, {\it Phys. Rev. Lett.\/} {\bf 98},
  010504 (2007).

\bibitem{Ursin2007}
R.~Ursin, {\it et~al.\/}, {\it Nat. Phys.\/} {\bf 3}, 481 (2007).

\bibitem{Majumdar2015}
R.~E. Meyers, {\it Advanced Free Space Optics (FSO): A Systems Approach\/}
  (Springer, 2015), chap. Free-Space and Atmospheric Quantum Communications.

\bibitem{Lorenz2004}
S.~Lorenz, N.~Korolkova, G.~Leuchs, {\it Appl. Phys. B\/} {\bf 79}, 273 (2004).

\bibitem{Lorenz2006}
S.~Lorenz, {\it et~al.\/}, {\it Phys. Rev. A\/} {\bf 74}, 042326 (2006).

\bibitem{Elser2009}
D.~Elser, {\it et~al.\/}, {\it New J. Phys.\/} {\bf 11}, 045014 (2009).

\bibitem{Peuntinger2014}
C.~Peuntinger, {\it et~al.\/}, {\it Phys. Rev. Lett.\/} {\bf 113}, 060502
  (2014).

\bibitem{Heim2014}
B.~Heim, {\it et~al.\/}, {\it New J. Phys.\/} {\bf 16}, 113018 (2014).

\bibitem{Elser2015}
D.~Elser, {\it et~al.\/}, {\it Proc. ICSOS\/} (IEEE, 2015).

\bibitem{Smutny2009}
B.~Smutny, {\it et~al.\/}, {\it Proc. SPIE 7199\/}, H.~Hemmati, ed. (2009), p.
  719906.

\bibitem{Heine2015}
F.~Heine, {\it et~al.\/}, {\it Proc. SPIE 9354\/}, H.~Hemmati, D.~M. Boroson,
  eds. (2015), p. 93540G.

\bibitem{Zech2015}
H.~Zech, {\it et~al.\/}, {\it Proc. SPIE 9647\/}, E.~M. Carapezza, P.~G.
  Datskos, C.~Tsamis, L.~Laycock, H.~J. White, eds. (2015), p. 96470J.

\bibitem{Heisenberg1927}
W.~Heisenberg, {\it Z. Phys.\/} {\bf 43}, 172 (1927).

\bibitem{Fischer2015}
E.~Fischer, {\it et~al.\/}, {\it Proc. ICSOS\/} (IEEE, 2015).

\bibitem{Glauber1963}
R.~J. Glauber, {\it Physical Review\/} {\bf 131}, 2766 (1963).

\bibitem{Bennett1992}
C.~H. Bennett, {\it Phys. Rev. Lett.\/} {\bf 68}, 3121 (1992).

\bibitem{Killoran2012}
N.~Killoran, M.~Hosseini, B.~C. Buchler, P.~K. Lam, N.~L\"utkenhaus, {\it Phys.
  Rev. A\/} {\bf 86}, 022331 (2012).

\bibitem{Yuen1978}
H.~P. Yuen, J.~H. Shapiro, {\it Coherence and Quantum Optics IV\/}, L.~Mandel,
  E.~Wolf, eds. (Plenum Press, New York, 1978), p. 719.

\bibitem{Ralph1999}
T.~C. Ralph, {\it Phys. Rev. A\/} {\bf 61}, 010303 (1999).

\bibitem{Silberhorn2002}
C.~Silberhorn, T.~C. Ralph, N.~L{\"{u}}tkenhaus, G.~Leuchs, {\it Phys. Rev.
  Lett.\/} {\bf 89}, 167901 (2002).

\bibitem{Vogel2016}
D.~Vasylyev, A.~A. Semenov, W.~Vogel, {\it Phys. Rev. Lett.\/} {\bf 117},
  090501 (2016).

\bibitem{Note1}
The measurement of the $P$ quadrature will be subject of future studies.

\bibitem{Inoue2002}
K.~Inoue, E.~Waks, Y.~Yamamoto, {\it Phys. Rev. Lett.\/} {\bf 89}, 037902
  (2002).

\bibitem{Wang2005}
X.-B. Wang, {\it Phys. Rev. Lett.\/} {\bf 94}, 230503 (2005).

\bibitem{Ma2005}
X.~Ma, B.~Qi, Y.~Zhao, H.-K. Lo, {\it Phys. Rev. A\/} {\bf 72}, 012326 (2005).

\bibitem{Lo2005}
H.-K. Lo, X.~Ma, K.~Chen, {\it Physical Review Letters\/} {\bf 94}, 230504
  (2005).

\bibitem{Hwang2003}
W.-Y. Hwang, {\it Phys. Rev. Lett.\/} {\bf 91}, 057901 (2003).

\bibitem{Shibata2014}
H.~Shibata, T.~Honjo, K.~Shimizu, {\it Opt. Lett.\/} {\bf 39}, 5078 (2014).

\bibitem{Korzh2015}
B.~Korzh, {\it et~al.\/}, {\it Nat. Photon.\/} {\bf 9}, 163 (2015).

\bibitem{Bourgoin2015}
J.-P. Bourgoin, {\it et~al.\/}, {\it Phys. Rev. A\/} {\bf 92}, 052339 (2015).

\bibitem{Jouguet2013}
P.~Jouguet, S.~Kunz-Jacques, A.~Leverrier, P.~Grangier, E.~Diamanti, {\it Nat.
  Photon.\/} {\bf 7}, 378 (2013).

\bibitem{Walk2013}
N.~Walk, T.~C. Ralph, T.~Symul, P.~K. Lam, {\it Phys. Rev. A\/} {\bf 87},
  020303 (2013).

\bibitem{Wilde2013}
M.~M. Wilde, {\it Quantum Information Theory\/} (Cambridge University Press,
  2013).

\bibitem{Endo2016}
H.~Endo, {\it et~al.\/}, {\it Opt. Express\/} {\bf 24}, 8940 (2016).

\end{thebibliography}

%%%%%%%%%%%%%%%%%%%%%%%%%
%SUPPLEMENTARY
%%%%%%%%%%%%%%%%%%%%%%%%%

\newpage
\renewcommand\thefigure{S\arabic{figure}}    
\setcounter{figure}{0}

\section{Supplementary Material}
\subsection{Detailed description of signal analysis}
In the following, we describe the data analysis of the measurement leading to the result shown in Fig. 3 (main article) in detail. The investigated data was acquired on January 27, 2016 during an instant when the satellite was in a mode sending amplitude modulated binary phase shift keying (BPSK) signals. An oscilloscope (Teledyne LeCroy WaveRunner 640Zi, 4 GHz analogue bandwidth) was connected to the minus signal of the homodyne detector in the TAOGS (see Fig. 2 in the main article). We analyze an oscilloscope trace with a length of 1.25 ms corresponding to 50 million recorded samples with a sampling rate of 40 GS/s.

\begin{figure}[htb]
        \includegraphics[width=200px]{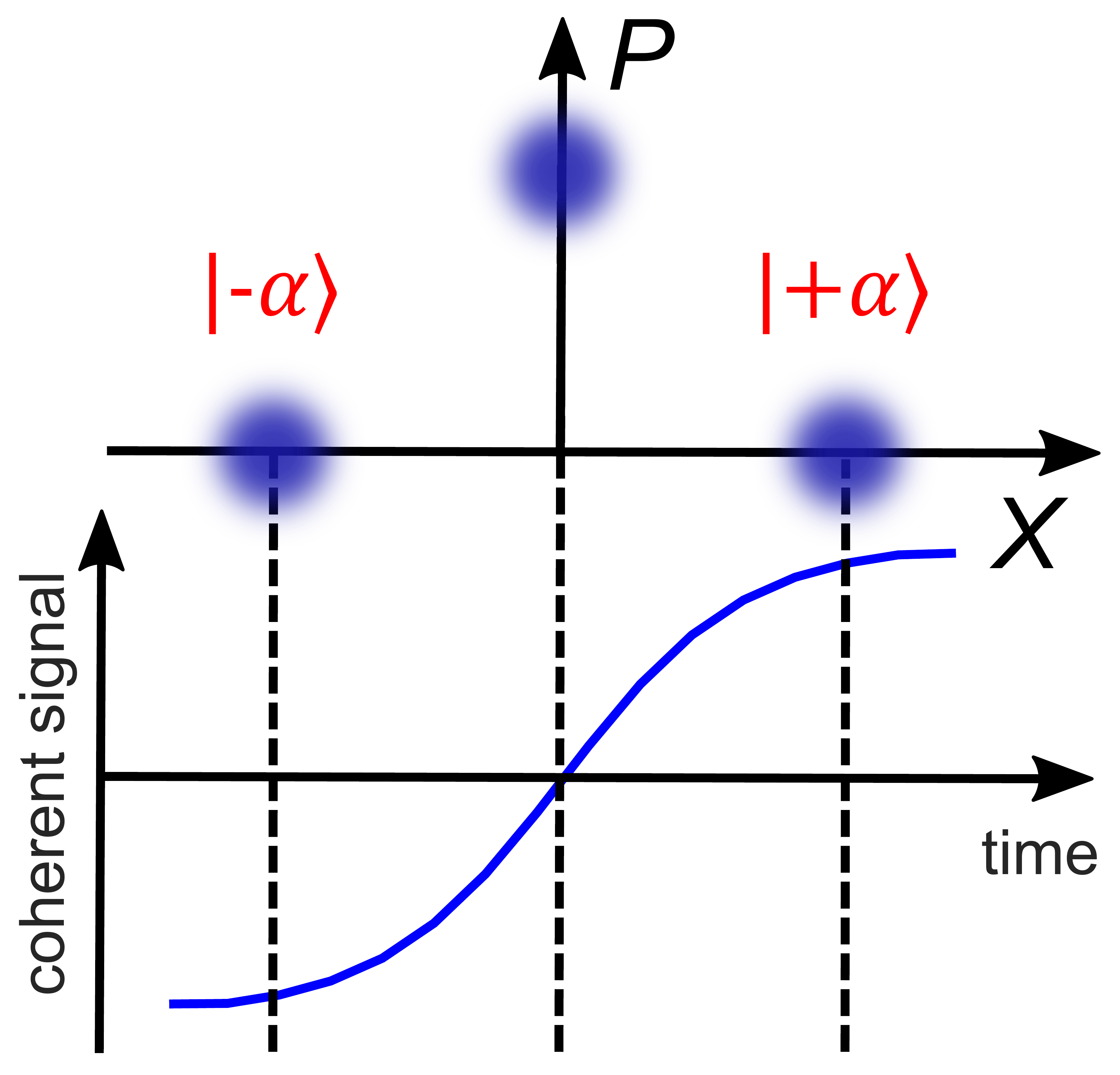}
        \caption{\label{figS1}Transient of binary phase shift keying between with the phase being switched from $\pi$ (for the state $|\mathord{-}\alpha\rangle$) to 0 (for the state $|\mathord{+}\alpha\rangle$). X and P are the electric field quadratures.}
\end{figure}

The first step of our analysis is to digitally recover the clock. This is necessary since our sampling rate of 40 GS/s is asynchronous to the BPSK signal frequency of 2.8125 GHz. Therefore, the duration of one signal pulse corresponds to the duration of approximately 14.2 oscilloscope sampling periods. We consider one sample per signal pulse for the following reason: At our receiver, we measure the projection onto the $X$ quadrature of a BPSK signal created by the phase modulator. Thus, as shown in Fig. \ref{figS1}, we resolve the finite switching time of the phase from 0 (for the state $|\mathord{+}\alpha\rangle$) to $\pi$ (for the state $|\mathord{-}\alpha\rangle$). As we do not acquire the $P$ quadrature simultaneously, the received amplitude $\alpha$ can be determined for the cases where we know that $\langle P \rangle=0$. This condition is fulfilled between pulse transients with the phase being either 0 or $\pi$. For that reason, we restrict ourselves to the samples in the pulse center. Additionally, the analogue bandwidth of the oscilloscope is 4 GHz and consequently neighboring samples taken at 40 GS/s are correlated. Therefore, the quantum state considered is defined by the time duration of one sample per signal pulse at the pulse center.

\begin{figure}[htb]
        \includegraphics[width=250px]{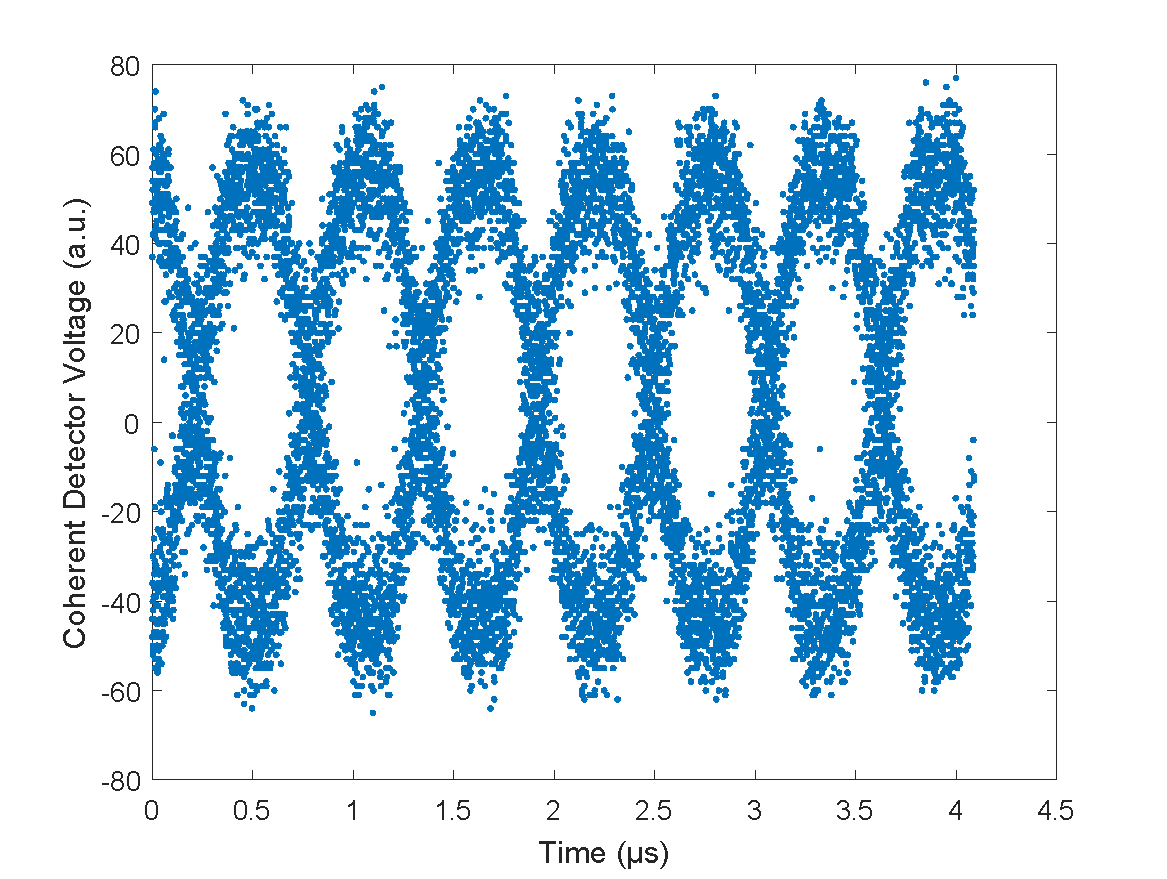}
        \caption{\label{figS2}Amplitude (sine wave) modulated BPSK signal samples at the signal frequency of 2.8125 GHz}
\end{figure}

The goal of the clock recovery is to determine the pulse centers as precisely as possible throughout the whole trace. The two parameters that need to be recovered are the offset between the BPSK signal clock and the oscilloscope clock and the frequency ratio between the clocks (i.e. the number of oscilloscope samples per signal pulse). We vary these two parameters successively to estimate the pulse centers and optimize for the sum of the absolute values of the respective coherent signals. Thereby we find an offset of 5.2380(5) samples and a frequency ratio of 14.2222731533(3) samples corresponding to a signal frequency of 2.81249(1) GHz. Note that the precision of the measured absolute frequency is limited by the clock accuracy of 4 ppm of our oscilloscope. 
According to this recovery, we take the samples closest to the values 5.2380(5) + n $\cdot$ 14.2222731533(3) (n=0, 1, ..., 7031223). A fraction of the resulting signal trace is shown in Fig. \ref{figS2}.

In the next step, we analyze the $X$ quadrature distribution for the different signal amplitudes. During BPSK operation, the amplitudes are modulated with a MHz sine wave. The amplitude modulation period repeats every 1600 signal states, corresponding to a duration of 0.568890(3) $\mu$s.

\begin{figure}[htb]
        \includegraphics[width=250px]{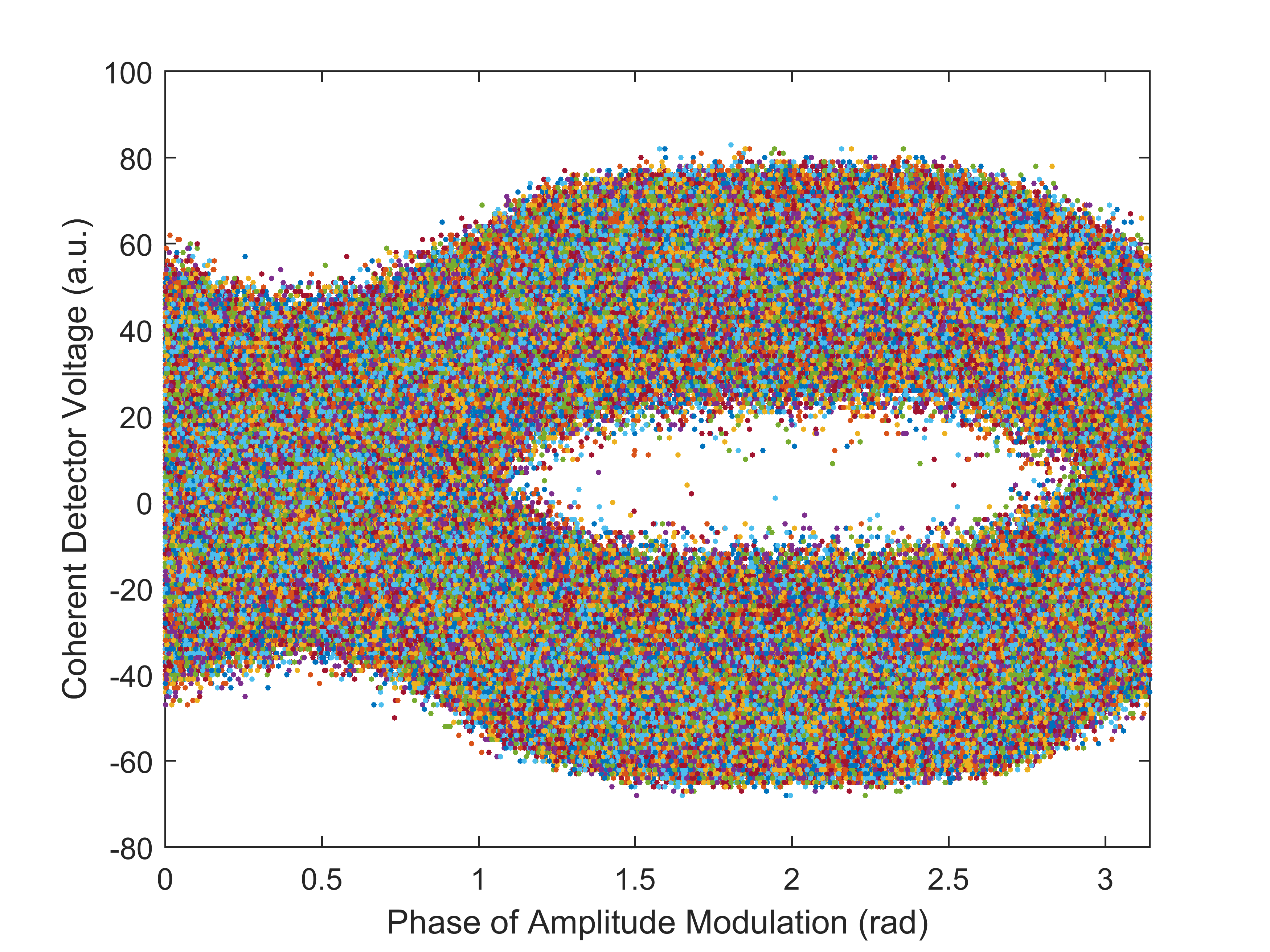}
        \caption{\label{figS3}The amplitude modulator generates a slow envelope for BPSK signals. Half-periods of this envelope are superimposed for further statistical data analysis. As the receiver becomes non-linear for $\alpha \geq 2$ (saturation at peaks), we restrict our analysis to the amplitudes smaller than 2 (left part of the trace).}
\end{figure}

In order to determine the coherent amplitudes $\alpha$ and excess noise (noise variance above the vacuum quantum uncertainty), the repeating amplitude modulated periods are superimposed for the duration of 1.25 ms for which the channel transmission is measured to be stationary. This duration corresponds to 2197 period repetitions. The resulting trace is shown in Fig. \ref{figS3}.

The superimposed traces of Fig. \ref{figS3} are binned into 50 segments of phase 2$\pi$/100, corresponding to a duration of 11.3778(1) ns for each segment. This binning yields 70304 data points for each segment. 

\begin{figure}[htb]
        \includegraphics[width=250px]{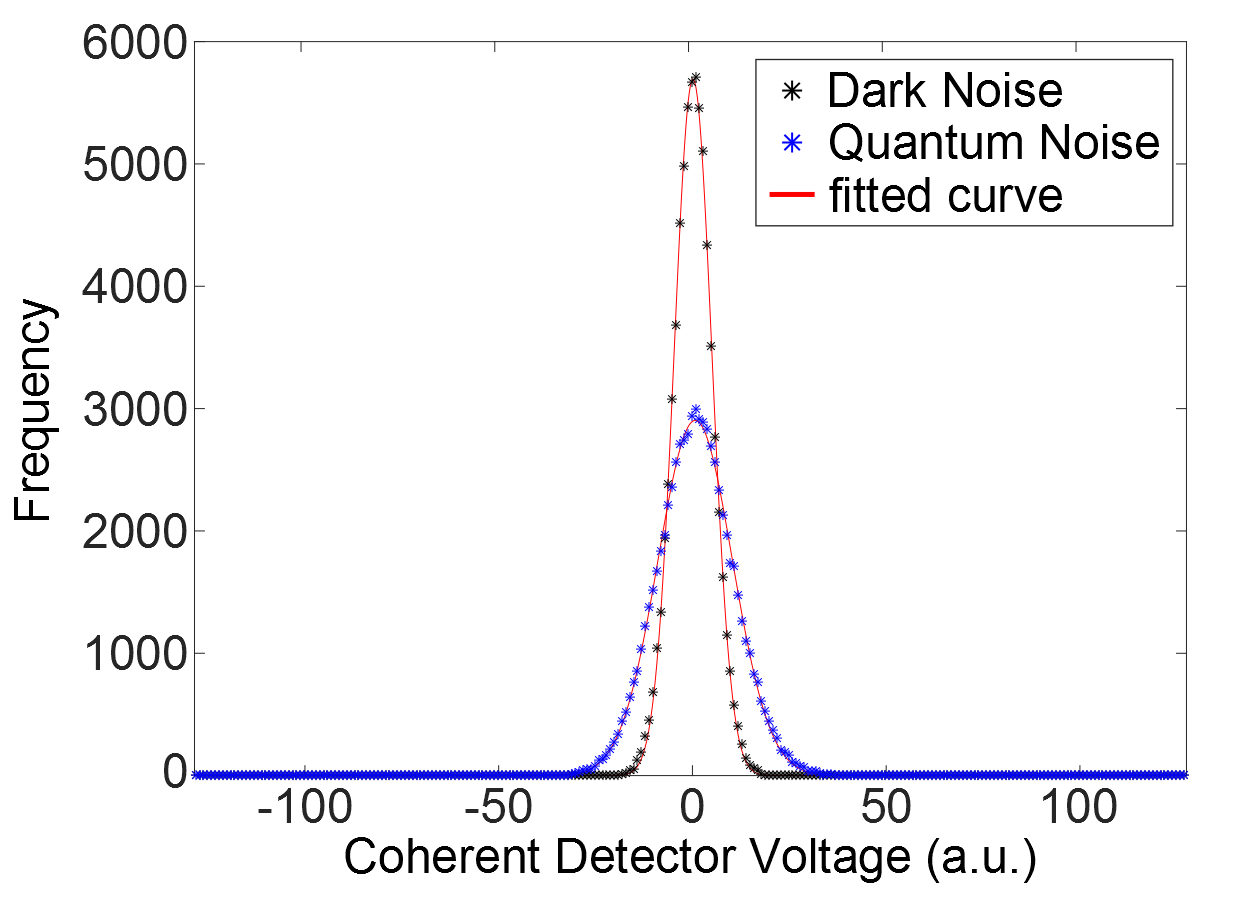}
        \caption{Homodyne measurement histograms of vacuum quantum noise and electronic dark noise. The clearance between the recorded vacuum quantum noise and the electronic noise of the detector is 6 dB.}
        \label{figS4}
\end{figure}

For each segment, histograms are generated. For the BPSK signal we fit a double Gaussian function of the form 
\[ a_{1}  \cdot \exp\left[-\left(\frac{x-b_{1}}{c_{12}}\right)^2\right] + a_{2} \cdot \exp\left[-\left(\frac{x-b_{2}}{c_{12}}\right)^2\right]. \]
Note that we assume here that the variance of the two signal states $|\mathord{+}\alpha\rangle$ and $|\mathord{-}\alpha\rangle$ is equal. Vacuum quantum noise and dark noise traces which have been recorded beforehand are processed equally with a single Gaussian fit function $a_{1} \cdot \exp(-((x-b_{1})/c_{1})^2)$ , see Fig. \ref{figS4}. 

From the fit parameters $b_{1}$ and $b_{2}$ of the double Gaussian fit, the coherent state amplitudes $\alpha$ of the signals are determined by normalizing the vacuum quantum noise variance to 1. As the gain in the consecutive electronic amplification stage after the coherent receiver becomes non-linear for $\alpha \geq 2$, we restrict our analysis to the 13 segments where the received amplitude was smaller than 2.

For these amplitudes, the fitted curves agree well with the measured data. Three exemplary fits are shown in Fig. 3 of the main article. The quality or goodness of fit given as the coefficient of determination $R^2$ is in average 0.999985 for the considered amplitudes.

Finally, we determine the excess noise above the quantum uncertainty of the vacuum state for the different amplitudes. For each of the considered amplitude modulation segments, the fitting yields variances of dark noise, vacuum quantum noise and signal states, as well as the 95\% confidence intervals. For electronic dark noise and vacuum quantum noise, which are not amplitude modulated, the mean value of all samples is taken. Dark noise is subtracted from vacuum quantum noise and signal noise (clearance between vacuum quantum noise and dark noise is given in the following section).
The excess noise or excess variance $E$ of an observable $X$ for a signal state is calculated by comparing its variance to the variance of a coherent vacuum state (quantum uncertainty):
\begin{equation}
E(X) = \frac{\Delta^2 X(\mathrm{signal})}{\Delta^2 X(\mathrm{vacuum})} - 1
\end{equation}
This calculation results in an excess noise value for each of the 13 measured amplitude bins. 

To estimate the error on our measured excess noise values we use the 95\% confidence intervals of the relevant fit parameters for dark noise, vacuum quantum noise and signal variance. For the excess noise error, we perform a worst-case estimation, i.e. we calculate the largest and smallest possible value for $E$, taking into account the confidence intervals. The same procedure is performed to obtain error bounds for the measured amplitudes $\alpha$. In addition we take into account the error arising from the binning procedure described above, which collects slightly different amplitudes in one bin. To this end, we perform a sinusoidal fit to the modulated amplitudes to estimate the binning error $\lbrack \mathrm{S1} \rbrack$. The resulting measured amplitudes and excess noise values together with the respective estimated error bars can be found in Fig. 3 of the main article. 

\subsection{Detector characterization}
We verify the linearity of the homodyne detection system by step-wise attenuation of the power of the local oscillator and verifying that the variance of the vacuum state scales linearly. 

\begin{figure}
        \includegraphics[width=250px]{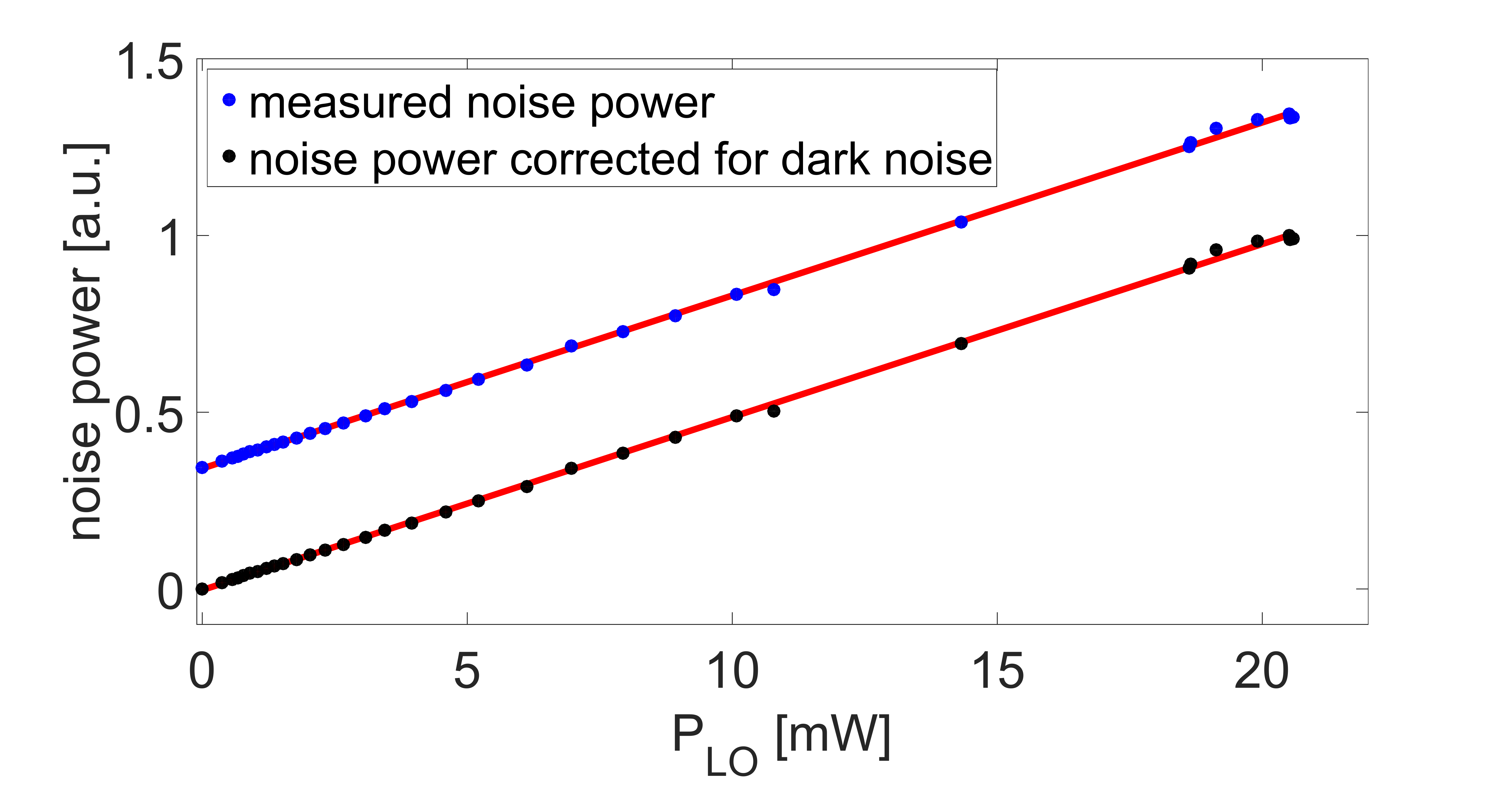}
        \caption{Linear scaling of vacuum quantum noise measurements with local oscillator power}
        \label{figS5}
\end{figure}

The measured power values were used to calibrate the sum of the DC outputs of our detectors in mW. These DC output values are included in the telemetry data, recorded by the TAOGS at a frequency of 5 kHz. The noise power was calculated for each local oscillator power from the recorded traces. We use the same processing routine for the vacuum state and for the satellite signal states, except that a single Gaussian is fitted to the vacuum state and a double Gaussian to the BPSK signal states (see above). From the Gaussian fits we extract the noise variances. In Fig. \ref{figS5} the measured noise power as well as the noise power corrected for dark noise is plotted versus local oscillator power.

A linear fit shows that the scaling of the noise power is linear. We find a clearance of up to 6\,dB of the quantum vacuum noise power measured with the respective local oscillator power above the detector's electronic noise. During satellite communication, we use the telemetry data provided by the TAOGS to determine to the corresponding vacuum variance for each signal measurement run.

\subsection{Amplifier noise behavior}
The BPSK signals in the LCT are amplified before being sent towards the ground station. The amplifier exhibits a noise figure of $\mathrm{NF} = \frac{\mathrm{SNR_{out}}}{\mathrm{SNR_{in}}} = 6.4$\,dB, which is the ratio between the signal-to-noise ratio of the input signal and the signal-to-noise ratio of the amplified signal. Together with the amplifier gain of 27\,dB, this translates into thermal quantum states with a quadrature variance of about 33\,dB above the quantum uncertainty of the vacuum state. See main article for discussion.

\subsection{Virtual aperture concept}
Excess noise is defined as noise above the quantum uncertainty of the vacuum state. In the following, we will bound the amount of excess noise that Earth’s atmosphere could have added to the quantum states from space. The signal states at the satellite are well above the quantum uncertainty of the vacuum state (see section above). This sender noise would at first glance impair the precision of a noise characterization. On the other hand, we can benefit here from the fact that an aperture, which is smaller than the beam diameter, sees reduced excess noise. As a gedankenexperiment, we imagine a virtual aperture in the exosphere at an altitude of 1000 km above ground (see Fig. 4 in the main article). Between geostationary Earth orbit (GEO) and exosphere $\lbrack \mathrm{S2} \rbrack$, light propagates quasi in vacuum which we assume not to add noise for this consideration. 
In the following we give an estimate of the diffraction losses:
In the far field the intensity distribution of our beam is given as an Airy pattern due to diffraction at our sending aperture. In good approximation we can assume that the intensity distribution at the center of the Airy pattern is Gaussian. The fraction of the detected power within the virtual aperture can thus be estimated by integration over the intensity ratio

\begin{equation}
\frac{I(r,z)}{I_{0}} = \left(\frac{w_{0}}{w(z)}\right)^2 e^{-\frac{2r^2}{w(z)^2}},
\end{equation}
where $w_{0} = 6 \,\mathrm{cm}\,(z_{0} = 10.6 \,\mathrm{km})$ is the beam waist, $w(z) = w_{0} \sqrt{1+(\frac{z}{z_{0}})^2}$ the spot radius at a distance $z = 38600 \,\mathrm{km}$ from the satellite and $r$ the radial distance to the optical axis. Integration over r yields
\begin{equation}
\frac{P(r_{0}, z)}{P_{0}} = 1 - e^{-\frac{2r_{0}^2}{w(z)^2}} = 61\,\mathrm{dB}
\end{equation}
of loss between the satellite and the virtual aperture $(r_0 = 13.5 \,\mathrm{cm})$. 
Scaling the initial excess noise of 33\,dB with this loss results in a variance merely -28\,dB above the quantum uncertainty of the vacuum state. In linear units, this corresponds to a variance of 0.001 units above the quantum uncertainty of the vacuum state. Thus, for practical purposes, the states at the virtual aperture can be considered to be quantum noise limited. Therefore, all excess noise at the aperture of the ground station must have been added by the atmosphere. Thus, all detected excess noise originates either from the atmosphere and/or from technical noise inside the TAOGS. In the following, we conservatively assess the noise contribution of the atmosphere by assuming that the receiving station operates noiselessly. We scale the measured excess noise at the homodyne detector $(0.01\pm0.03)$ (see Fig. 3 in the main article) with 16\,dB of technical losses in the station (including detector efficiency). This yields $0.4\pm1.2$ of excess noise at the TAOGS aperture. Considering the virtual aperture concept, we additionally scale this value with atmospheric losses (absorption, scintillation and scattering) of 3\,dB and obtain $0.8\pm2.4$. This is the sought bound for atmospherically added excess noise.

~\\
$\lbrack \mathrm{S1} \rbrack$ V. C. Usenko, B. Heim, C. Peuntinger, C. Wittmann, C. Marquardt, G. Leuchs and R. Filip, "Entanglement of Gaussian states and the applicability to quantum key distribution over fading channels," New J. Phys. \textbf{14}, 093048 (2012).
~\\
$\lbrack \mathrm{S2} \rbrack$ European Cooperation for Space Standardization, ECSS-E-ST-10-04C, Space environment, Second issue, ESA-ESTEC, Requirements \& Standards Division, Noordwijk, The Netherlands, 15 November 2008.

\end{document}